# Fine structure and optical pumping of spins in individual semiconductor quantum dots

**Allan S Bracker,[1] Daniel Gammon[1] and Vladimir L Korenev[2]**

[1]Naval Research Laboratory, Washington, DC  20375 USA

[2]A. F. Ioffe Physico-Technical Institute, St. Petersburg, 194021 Russia

Email: bracker@bloch.nrl.navy.mil

**Abstract.** We review spin properties of semiconductor quantum dots and their effect on optical spectra. Photoluminescence and other types of spectroscopy are used to probe neutral and charged excitons in individual quantum dots with high spectral and spatial resolution.  Spectral fine structure and polarization reveal how quantum dot spins interact with each other and with their environment. By taking advantage of the selectivity of optical selection rules and spin relaxation, optical spin pumping of the ground state electron and nuclear spins is achieved.  Through such mechanisms, light can be used to process spins for use as a carrier of information.

## 1. Introduction

Not long after the first spectra of individual quantum dots (QDs) were reported [1,2,3,4], their spin fine structure and spin polarization became a focus of fundamental interest through the measurements of polarized spectral fine structure [5] and hyperfine structure (Overhauser shifts) [6].  On one hand, spin provided highly detailed information with which to test the analogy between QDs and real atoms [7].  QD spins were also regarded as a potential form of quantum information [8].  Spatial confinement inhibits some decoherence mechanisms and allows the application of localized electrical or optical interactions.  The techniques of single dot spectroscopy provide one way to manipulate and measure individual spins.  The physics of spins in QDs was reviewed extensively in 2003 [9], with emphasis on experimental results in GaAs interface QDs.  In the last five years, there have been remarkable advances in experimental techniques and in the physical understanding of QD spins and their optical spectra.  In this paper, we consider examples examples from our own work, emphasizing the importance of high spectral and spatial resolution that is achieved with single QD spectroscopy.

The problem of QD inhomogeneity was the original motivation for single dot spectroscopy, and it is worth reemphasizing this in the context of spin measurements. The width of an ensemble optical spectrum originates largely from statistical variability in the growth process, which produces a range of sizes, shapes, and compositions of QDs. Consider just how serious the problem is: A typical ensemble of InAs self-assembled QDs may have photoluminescence energies spread over a featureless peak with a width of 30 meV ["s" shell in Fig. 1(a)]. A single QD may contribute 10 or more very narrow peaks to this broad spectrum, corresponding to different charge and spin configurations [Fig. 1(d)]. To date, it is difficult to achieve energy homogeneity for an ensemble of InAs dots that is comparable to typical Coulomb energy shifts for different charge configurations (<10 meV). It is almost impossible to imagine approaching the smaller spin splittings (tens of µeV), let alone the homogeneous linewidth (1 µeV) of an individual optical transition. It is likely that growth-induced inhomogeneity will be with us for a long time to come, and researchers must design device architectures and physics experiments that account for this fact.

## 2. Single dot spectroscopy techniques

The ability to spatially and spectrally isolate an individual dot is the technique at the heart of the field. The principle is very simple—to shrink the measured part of a sample to such a small size that the spectral lines of individual QDs are well separated in energy. Several techniques have been used—metal shadow masks with apertures [10,11], etched mesas [3], and a purely optical approach using a small laser spot or small field of view [2,4,12]. A microscope lens in conjunction with a solid immersion lens has been used to focus a laser spot down to 250 nm for single dot spectroscopy [13]. Scanning near-field imaging spectroscopy has reached an order of magnitude smaller [14]. The aperture and mesa methods are achieved using electron beam lithography combined with other conventional fabrication techniques. These methods can perturb the sample—the metal aperture may strain the material or modify the optical field of incoming or outgoing light, while small mesas bring the QDs into the proximity of sample surfaces. The optical methods avoid these problems. One disadvantage of the optical methods is that it can be difficult to locate a particular QD the next day (or the next year) when the entire sample area is covered with dots.

In our laboratory, we use primarily the shadow mask technique [11], where small apertures in the mask allow light to pass in and out. The shadow masks are formed by patterning a negative photoresist with an e-beam writer. After development, resist pillars remain, onto which aluminum is evaporated (typically 120 nm thick). The process is completed by liftoff of the metal-covered resist in an appropriate solvent.

The choice of aperture size in a shadow mask is governed by the density of spectral lines for each dot and the QD spatial densities that can be achieved. As we have already noted, an individual QD exhibits a cluster of spectral lines corresponding to various charge configurations (e.g. neutral, extra hole, extra electron, etc.) and their spin fine structure. These lines are spread out over roughly 10 meV for InAs self-assembled QDs [15] [Fig. 1(d)] and 5 meV for GaAs interface dots [16] [Figs 2(a) and (c)]. In order to easily assign the different peaks, these clusters should be separated from each other by more than this energy. InAs dots can be grown over a wide range of densities ($10^7$-$10^{11}$ cm$^{-2}$), and aperture diameters of 1 μm are typical for much of our work on InAs QDs. GaAs dots are in the higher end of this density range, and the ensemble spectrum is spread over a narrower range of energies. When measuring GaAs dots in a 1 μm aperture, we usually focus on outliers at the low energy edge of a crowded spectrum. Alternatively, apertures as small as 100 nm can be used.

A second key technique in single spin spectroscopy is the use of a Schottky diode heterostructure to select the charge configuration of the QD [Figs 1(b) and 2(b)]. This approach was already used for quantum well work and proved particularly powerful for the study of individual dots [15,17]. Since their introduction into QD work, many researchers have adopted them where practical. Their great advantage becomes clear by examining QD photoluminescence spectra as a function of diode bias. Typical spectral maps are shown for individual GaAs QDs in Fig. 2(c) and InAs QDs in Fig. 1(d). A series of streaks correspond to different charge configurations, separated in energy because of the different Coulomb interactions for each configuration. By considering these characteristic energies in combination with the charging order and spin fine structure, one obtains an unambiguous assignment of nearly every feature in the spectrum.

A typical diode structure consists of a heavily doped buffer layer topped with an insulating region in which the dots are embedded. The dots themselves are typically 20-80 nm above the heavily doped region. The thickness of the insulating spacer affects the tunneling rate, and interaction between the dot and the electron reservoir can be a rich source of spin physics [18]. An AlGaAs "current blocking barrier" is sometimes included in the insulating region above the QDs. A semi-transparent contact (e.g. 5 nm titanium) and the aluminum shadow mask are deposited on top of the insulating region. At low electric fields (forward bias in an n-type diode), the confined states of the dot fall below the quasi-Fermi level, and an electron tunnels into the dot from the buffer. The bias at which this occurs depends not only on the single particle energy of the QD, but also on the Coulomb interaction of the electron with other charges already in the QD [19]. Dissociation of *e-h* pairs after optical excitation often produces minority carrier charged excitons [e.g. $X^+$ in Figs 1 and 2].

Many of the optical techniques used for spin measurements on single dots are the same as those used for ensembles and have been discussed in *Optical Orientation* [20] and elsewhere [21]. Photoluminescence detection is by far the most common technique for single QD spectroscopy, but as the field has advanced, other forms of linear and nonlinear spectroscopy have been added to the repertoire [See for example, Refs 22,23,24,25].

The challenge of resolving fine structure in ensemble spectra arises in part because we have chosen to use bandgap spectroscopies such as photoluminescence or transmission, where small splittings are masked by large inhomogeneous broadening. However, the situation is really more favorable than it first appears, because we can use magnetic fields to directly probe the small energy scales. This point is very clear in the concept of "level crossing spectroscopy," of which the Hanle effect is a special case. The idea is to bring two separated atomic energy levels together (or apart in the case of the Hanle effect) by using the Zeeman effect [Fig. 4(b)]. When the levels become degenerate to within their natural linewidth, a coherence is produced, and the polarization or direction of light emission changes. A magnetic field can also be used to change the polarization of an exciton from linear to circular [Fig. 3(b)], providing a characterization of *e-h* exchange splittings. Alternatively, time resolved techniques can be used to directly measure the frequencies associated with spin splittings. Although this review emphasizes the power of high spectral resolution, we briefly consider the magnetic field-based depolarization techniques in Section 4. These methods are powerful when applied to ensembles and can substantially enhance single dot spectroscopy as well.

Most of the work discussed in this review involves two classes of QDs—self-assembled InAs QDs and GaAs interface fluctuation QDs, and it is helpful to summarize several important differences between them. InAs QDs are significantly strained and typically smaller than the GaAs dots in their lateral dimension. The height of the confining potential is governed by InAs/GaAs band offsets, strain, and quantum confinement. For the InAs QDs in our work, 100-250 meV (shared between electron and holes) is typical. The QDs are usually pancake-shaped, with lateral dimensions of 10-20 nm and vertical dimensions of 2-10 nm. A ladder of several orbital states ("s, p, d, f...") is confined below the 2-dimensional InAs wetting layer continuum.

Typical radiative lifetimes are around a nanosecond. The GaAs QDs differ substantially in both morphology and electronic structure. They are formed by monolayer thickness fluctuations in a narrow (3-4 nm) GaAs quantum well. They are nearly unstrained, and the height of the potential in the growth direction is strong, governed by the GaAs-AlGaAs band offset and quantum confinement. Confinement within the plane of the quantum well is weak (~10 meV) and arises from the thickness fluctuations. The lateral dimensions of the GaAs QDs vary widely—from a few nanometers to a hundred nanometers or more at the extremes. When the lateral size is small, very few confined states may exist below the quantum well continuum. When the lateral size is large, it may exceed the exciton Bohr radius, and excited states may be separated by only a few meV. Radiative lifetimes are typically below 100 ps.

## 3. Spin fine structure

With single dot spectroscopy, we spectrally resolve, assign, and manipulate individual spin quantum states. In this section, we review the fine structure and polarization selection rules of QD spin states for the most-studied charge configurations—neutral and singly charged dots and their lowest excitonic states. We also consider excited single-particle orbital states and the effect of nuclear polarization on spectral fine structure.

*3.1 Neutral quantum dot*

When sharp atom-like peaks were first observed in spectra of individual GaAs QDs, it was believed that the peaks came from neutral excitons. In retrospect, this initial assignment was somewhat lucky, because it turns out to be quite easy to observe neutral, positive, and negative excitons all within the spectrum of one dot [16,19], even for undoped samples. However, with the benefit of spectral fine structure and photoluminescence polarization measurements, it was quickly shown that the initial assignment of a neutral exciton was indeed correct [5].

The exciton fine structure is shown in Fig. 3(a). The ground state of the optical transition is an empty QD, denoted by $|0\rangle$. The neutral exciton has four spin states. The details of their energy splittings are strongly influenced by the flat cylindrical disk shape of typical epitaxial QDs. With this symmetry, the four states are arranged into two doublets with nominal spin basis functions corresponding to the four spin pairings of one electron and one heavy hole. The upper doublet consists of excitons with antiparallel electron and hole spins ⇑↓ and ⇓↑, which have angular momentum projections m= ±1, while the lower energy pair has parallel spin combinations ⇑↑ and ⇓↓, with m= ±2, respectively. The two doublets are split by the axially symmetric part of the *e-h* exchange interaction $\delta_0$, which is of order 100 μev. In fact, the dots have lower symmetry—both due to the growth process and the underlying crystal structure [26,27], and this causes the individual doublets to split into linearly polarized mixtures of the circular basis functions [5], with anisotropic *e-h* exchange splittings $\delta_b$ and $\delta_d$.

The magnetic field dependence of these states in the Faraday geometry is shown in Fig. 3(b). As the field increases, the states evolve from linear to circular as the Zeeman splitting becomes larger than the exchange splittings. These energy levels are described by:

$$E_{\pm 1} = \tfrac{1}{2}(\delta_0 \pm \sqrt{\delta_b^2 + h_1^2}) \quad \text{and} \quad E_{\pm 2} = \tfrac{1}{2}(-\delta_0 \pm \sqrt{\delta_d^2 + h_2^2}), \tag{1}$$

where $h_n = \mu_B g_n B_{ext} + (-1)^n A \langle I_z \rangle$, and the g-factor is $g_n = g_z^{hh} - (-1)^n g_z^e$, with contributions from both the electron and heavy hole. The second term in $h_n$ describes the influence of polarized nuclei and is discussed further in Section 3.4 and Section 6.

From angular momentum selection rules, only the upper pair of states composed of m= ±1 basis functions have optically allowed ("bright") transitions to the ground state. The "dark" states (m= ±2) can be observed directly in a magnetic field tilted away from the z axis (Fig. 5). The bright doublet can be resolved in polarization-sensitive PL spectroscopy [5] or transmission spectroscopy [28].

*3.2 Singly charged quantum dot*

The doublet structure and linear polarization of a neutral quantum dot disappear when the dot becomes charged [Fig. 4(a)] [29,30]. This degeneracy is a result of the Kramers theorem for particles with half-integer spin. In contrast, the neutral exciton has integer spins, so its degeneracy can be lifted completely. The ground state energy levels of the charged quantum dot consist of a degenerate Zeeman doublet for the single carrier, which can be an electron or a hole. The lowest optically excited state is the charged exciton (or "trion"), with two carriers of one type in a spin-paired singlet and one unpaired spin of the other carrier type. The unpaired spin distinguishes the two Zeeman components of the charged exciton, and gives a g-factor different from that of the optical ground state, which is determined by the other carrier spin. The trion has no zero-field fine structure, because there is only one unpaired spin and no exchange interaction.

The ground and excited states are connected through circularly-polarized selection rules. Diagonal transitions are optically forbidden at zero magnetic field or when the magnetic field is oriented along the QD axis and light propagation direction. These selection rules change when the magnetic field is tilted [Fig. 5], with all transitions becoming allowed and linearly polarized. As described in Section 5, this latter fact proves useful in designing an efficient resonant optical pumping scheme.

Describing the excited states of charged QD excitons is not trivial. Some quantum dots have lateral dimensions much larger than the trion Bohr radius. This is true of some GaAs interface QDs and of InP islands in an InGaP matrix [31]. In these systems, as in quantum wells, triplet states may be unbound. In QDs with a deep potential and small dimensions (e.g. InAs self-assembled dots), carriers are strongly confined in all directions. Coulomb interactions become comparable to or smaller than the kinetic energy associated with size quantization. The term "exciton" is preserved in literature, although the electron and hole are bound together more by the QD potential than by the Coulomb interaction. The presence of multiple single particle orbitals in small deep QDs allows for singlet and triplet states.

The triplet fine structure for self-assembled InAs QDs has been a topic of considerable interest in the last several years [32,33,34], mostly because of its connection to negative photoluminescence polarization and to optical pumping. The lowest trion excited states have one electron each in "s" and "p" orbitals [Fig. 6(a)]. Singlet $X_{S^*}^-$ and triplet $X_T^-$ configurations are separated by the *e-e* axially symmetric exchange interaction ($\Delta_{ee}$), with the excited singlet expected to lie at higher energy by several meV. The three components of the triplet manifold are also split by a much smaller (hundreds of μeV) *e-h* axially symmetric exchange interaction ($\Delta_{eh}$) [35]. Two of the three triplet components have been observed experimentally [34,36]. The assignment of these peaks is based partly on the circular polarization of photoluminescence and is discussed further in Section 4.

*3.3. Quantum dot molecule*

Quantum dot molecules give additional insight into the same spin interactions that apply to single QDs, and they include an essential new element—the coherent tunneling of charges between the dots [37,38, 39,40]. To date, vertically stacked self-assembled InAs QDs [Fig. 7(a)] are the most widely studied system using optical spectroscopy. By applying an electric field, one or more carriers can tunnel between the two dots, with delocalized "molecular" orbitals forming in the intermediate (coupled) region. A simplified schematic of exciton energy levels is shown in Fig. 7(b). The tunneling anticrossing occurs where the levels for spatially direct (intradot) and indirect (interdot) excitons meet. Here, the two types of excitons differ only in the position of the hole, which is the tunneling carrier.

It is revealing to examine how spin interactions evolve in the anticrossing region. Fig. 7(c) shows the situation for the neutral exciton. As noted earlier, a neutral exciton has two classes of spin states, bright and dark, separated by the *e-h* exchange interaction $\delta_0$. However, $\delta_0$ depends strongly on the wavefunction overlap between the electron and hole. As we follow each eigenstate through the anticrossing, the hole wavefunction shifts from the left dot to the right dot, and we observe the steady disappearance of the exchange splitting as the two carriers are pulled apart.

For the molecular trion, both *e-h* and *h-h* exchange combine to give a more complex total interaction. When all three carriers reside in the same dot, the spectrum resembles that of a singlet trion with no exchange splitting. However, when one hole is pulled into the other dot, the exchange splitting $\delta_0$ appears for the remaining electron and hole. The evolution of these states through the anticrossing is particularly illuminating. Both singlet and triplet states exist for the two-hole configuration. The triplet states pass through the anticrossing region, because the Pauli principle forbids unpaired spins to be in the same dot, and tunneling cannot occur. The two triplets with parallel hole spins (dotted lines) are completely unaffected by the anticrossing. However, the singlets anticross, and the triplet with two-hole spin projection m=0 mixes with the singlets in this region to produce the slight "wiggle" observed in the data. This wiggle comes about through the electron-hole exchange interaction that mixes the singlets and this one triplet.

*3.4 Effect of hyperfine interaction on electron spin level structure*

Under circularly polarized excitation, the nuclear spins in a QD can become polarized, and the resulting nuclear polarization modifies the electron spin energies. These processes result from the contact hyperfine interaction between electron and nuclear spins

$$\hat{H}_{hf} = \frac{v_0}{2} \sum_j A^j \left| \psi(\vec{r}_j) \right|^2 \left( \hat{I}_z^j \hat{\sigma}_z^e + \frac{\hat{I}_+^j \hat{\sigma}_-^e + \hat{I}_-^j \hat{\sigma}_+^e}{2} \right), \quad (2)$$

where $v_0$ is the volume of the unit cell, $A^j$ is the hyperfine constant, $\psi$ is the electron wavefunction, and the $\hat{I}$ and $\hat{\sigma}$ are the nuclear and electron spin operators. Hole spins have a negligible hyperfine interaction due to the p-like Bloch functions, which have little electron density at the nucleus. The first term in parentheses in Eq. 2 is responsible for the change in electron spin energies, and when averaged over all nuclei, behaves like an effective magnetic field. The second term describes the flip-flop interaction between electron and nuclear spins that causes the nuclei to be polarized. The physics of dynamic nuclear polarization (DNP) is discussed in detail in Section 6.

Here, we look back at the experimental evidence for polarized nuclei—the "Overhauser shifts" of the PL spectral line of an exciton in an individual QD [6]. The peak is measured for both circular polarizations of the exciting laser as a function of the applied magnetic field [shown for $\sigma^+$ in Fig. 8(a)], and the peak energies are plotted in Fig. 8(b). The energy averaged over the two polarizations is shown in Fig. 8(c). This characteristic quadratic dependence of the diamagnetic shift is polarization-independent, and shows that the orbital part of the electron wavefunction is not influenced by the polarized nuclei.

The Zeeman splitting [Fig. 8(d)] clearly shows the Overhauser shifts caused by nuclear polarization. Depending on whether the circular polarization of the laser is parallel or antiparallel to the applied magnetic field, the total splitting is larger or smaller than the intrinsic Zeeman splitting [Figs 8(a) and (b)]. (The splittings in Fig. 8(d) show a "dip" around $B_{ext}=0$. The origin of this dip is discussed further in Section 6.3.) The highest values of the Overhauser shift measured in our laboratory for GaAs interface QDs correspond to nuclear polarizations around 60% [41,42].

Nuclear magnetic resonance of the single QD gives further proof that the measured peak shifts are caused by polarized nuclei [43,44]. With a radiofrequency field tuned to the resonant energy of gallium or arsenic nuclei, the peak shift is nearly eliminated [Fig. 9]. The sensitivity of the measurement is extremely high—an NMR signal from less than $10^5$ nuclei is easily detected. Three properties make this possible—the high degree of DNP in QDs, the spectral resolution available with single QD spectroscopy, and the efficiency of photoluminescence detection.

## 4. Polarization spectroscopy

Photoluminescence polarization complements the information provided by fine structure energy splittings, and it provides a window into spin relaxation dynamics. The linear polarization of an exciton in a neutral dot is seen clearly in the PL. Linearly polarized light excites one of the bright doublet components and produces optical alignment of the exciton [5], which is maximal when the light is polarized parallel to the exciton transition dipole. Thus polarized PL directly monitors the reduced symmetry of the dot.

Measuring exciton polarization as a function of applied magnetic field provides a way to probe the small anisotropic exchange splittings ($\delta_b$ and $\delta_d$) of a quantum dot exciton even without high spectral resolution. In the same spirit as the Hanle effect, this technique can easily be applied to ensembles. A Faraday magnetic field overwhelms the asymmetries in the QD that produce the exchange splittings, leading to a change from linear to circular polarization as the Zeeman energy becomes larger than the anisotropic exchange splitting [5,45,31,46]. Fig. 3(b) shows this schematically for the anisotropic exchange of bright excitons. Although not strictly a "level crossing spectroscopy" (the levels anticross), this is an example where a magnetic effect is used to probe very small energy scales, without the need for high spectral resolution.

The PL of a singly charged QD is circularly polarized due to the degeneracy of its spin states. For resonant excitation of an s-shell singlet trion, the situation is very simple: $\sigma^+$ light excites trions $\downarrow\uparrow\Uparrow$ (or $\downarrow\Uparrow\Downarrow$) and $\sigma^+$ PL is emitted. (Some subtle but important exceptions are considered further in Section 5.2.) However, most PL measurements to date have used excitation into higher discrete states or continuum states ("nonresonant" excitation). In this case, even the coarse behavior is nontrivial to understand, because it reflects the physics of relaxation. One of the most intriguing observations is the phenomenon of negative PL polarization, which has been recognized in QD trions for more than a decade [31]. It has since been observed in a number of material systems, discussed below. Similar physics also results in reduced (although still positive) PL polarization in quantum wells [47].

For shallow or large QDs, s-shell trions form when the electron spins are antiparallel. Trion triplet states are probably weakly bound or unbound. The polarization behavior for this case was modeled for large InP islands [31] and later for GaAs interface QDs [42]. The incident $\sigma^+$-light creates free bright excitons ($\downarrow\Uparrow$), and usually the hole spins depolarize rapidly, while electron spins are more stable [48], leading to a mixture of bright and dark ($\Downarrow\downarrow$) excitons. Binding with the QD resident electron $\uparrow$ creates the singlet trions $\uparrow\downarrow\Uparrow$ ($\sigma^+$ PL) and $\uparrow\downarrow\Downarrow$ ($\sigma^-$ PL), respectively. In this simple scenario [9], the degree $\rho_c$ of circular polarization of trion PL is

$$\rho_c = \left( N_b \frac{P_c^b + P_s}{1 + P_c^b P_s} + N_d \frac{P_c^d - P_s}{1 - P_c^d P_s} \right) \Big/ (N_b + N_d), \qquad (3)$$

where the polarization (and density) of bright and dark excitons are $P_c^b$ ($N_b$) and $P_c^d$ ($N_d$), respectively, and the resident electron polarization is $P_s$. With the spin relaxation conditions discussed above, $P_c^b > 0$ and $P_c^d < 0$ under $\sigma^+$ excitation. Depending on the relative amounts of bright and dark excitons, the PL may have either sign, even when the resident electron is Bracker, et al., Fine structure and optical pumping of spins...

unpolarized ($P_s = 0$). For example, if the capture time of excitons is long, then bright excitons disappear through recombination before they can form trions [Fig. 10(b)], and $\rho_c < 0$ [31, 42]. Negative polarization can also result from fast depolarization of bright excitons ($P_c^b \approx 0$) [9]. In this situation, $\rho_c < 0$ even if the number of bright excitons is much larger than the number of dark, i.e. $N_b \gg N_d$.

Optical pumping of the resident electron spins ($P_s < 0$) drastically enhances the negative polarization signal [31, 42] through blocking of trion formation from the $\downarrow\Uparrow$ bright excitons produced by $\sigma^+$-light. This mechanism is discussed in Section 5.

Measuring the Hanle effect for trion PL polarization provides a way to observe long lifetimes. It is also a sensitive optically-detected method for probing the properties of resident electrons. The polarization of trion PL is measured as the magnetic field is varied in the Voigt geometry. In GaAs interface QDs, as in quantum wells, the analysis in simplified by the negligible value of the in-plane hole g-factor, so that only electrons are probed. The Hanle effect can be viewed as a "zero-field" level crossing spectroscopy between two Zeeman levels [Fig. 4(b)], where the levels are degenerate for all dots under the same condition (i.e. B=0). The peak width is governed by the inhomogeneous spin dephasing time $T_2^*$. The Hanle effect has been used to measure resident electron spin lifetimes as long as 300 ns in bulk GaAs [49], and 16 ns in single GaAs QDs [42]. Although these measurements are detected through photoluminescence, they are not limited by radiative linewidths, inhomogeneous broadening of the PL spectrum, or spectrometer resolution.

Fig. 4(c) shows the Hanle effect for a negative trion, positive trion, and neutral exciton in an individual QD. The neutral exciton remains linearly polarized in the magnetic field, so no circular polarization is detected. The Hanle linewidth for the electron in the positive trion is limited by the fast recombination time of the trion, so the peak is broad. In contrast, the negative trion shows a narrow downward-pointing peak. This sharp peak can be explained by the presence of polarized resident electrons with $T_2^* = 16$ ns. The negative PL polarization can be explained by the dark exciton model described above and demonstrates that optical pumping significantly enhances the negative polarization, as discussed in Section 5.

Negative PL polarization in InAs self-assembled QDs provides insight into the fine structure of trion excited states, where one electron is in a p-shell. This fine structure is observed directly in the PL excitation spectrum of a single QD, which shows a doublet with oppositely polarized peaks [34,36]. The peak with predominantly $\sigma^+$ PL polarization originates from resonant excitation to the middle state (*j=+3/2*) of the trion triplet on the left of Fig. 6(a). This state has zero projection of the electron angular momentum and relaxes quickly to the s-shell singlet trion and recombines. This rapid energy relaxation channel is not available for the $j=+1/2$ triplet trion, due to the Pauli principle. Instead, a combination of anisotropic *e-e* and *e-h* exchange interactions (dotted arrows) induces a simultaneous spin-flip of the p-shell electron and the hole, which allows relaxation to the ↑↓⇓ s-shell trion [32, 33,34]. Because the hole spin has flipped in this case, the PL polarization is negative.

Reports of negative PL polarization in InAs QDs under nonresonant excitation are quite common [32,33,50,51,52,53]. The situation is complex, and several of the proposed mechanisms may contribute. Holes and electrons in the wetting layer may be captured sequentially or as an exciton. As with GaAs or InP dots, the capture cross-section for electrons in InAs dots may be different for parallel and antiparallel electron spins, in which case, the kinetics of trion formation will occur according to Refs [31, 42] and Eq.(3). Once all three particles are within the QD, polarization reversal within the p-shell triplet manifold must be considered. In ensemble measurements, an analysis of negative PL polarization must also account for excitons containing more than one resident electron [33,50].

## 5. Optical pumping of resident electron spins

The term *optical pumping* originates with atomic physics experiments carried out nearly 60 years ago [54,55], and in recent years it has become common in the field of quantum dot physics. In this field, it is often associated with initialization of spin states for quantum information. In that context, it usually applies to polarized resident carrier spins that persist after short-lived excitonic species have recombined. In this section we consider optical pumping of electron spins, and in Sec. 6, we consider nuclear spins. In QDs, it is useful to distinguish between two classes of optical pumping, loosely called *resonant* and *nonresonant*, which differ in whether a discrete or continuum state is excited by the laser.

*5.1 Nonresonant optical pumping*

Nonresonant pumping involves excitation to a continuum such as the QD wetting layer, followed by relaxation and capture of carriers [Fig. 10(a)]. It has a long history that includes the well-studied optical orientation of donor electron spins in bulk semiconductors [56] and quantum wells [57]. Continuum excitation works well for an ensemble of dots, because a laser can place polarized carriers into a large number of dots over a wide area, and optical pumping signals can be quite large. However, high pumping fidelity is hard to achieve, because spin relaxation in the continuum tends to repopulate both resident spin states. Nonresonant optical pumping has been studied in a variety of materials in QD ensembles [31,32,52,53] and individual dots [42,58,59].

In Ref. [42], individual GaAs QDs were studied. The sample was excited with circularly polarized light tuned to the energy of the heavy hole quantum well continuum above the studied QD. The PL polarization of the positive and negative trions and neutral exciton were measured as a function of applied bias, laser intensity, laser polarization, and magnetic field. Strong evidence for optical pumping comes from the Hanle effect. The narrow Hanle peak for the negative trion in Fig. 4(c) corresponds to a lifetime of 16 ns. This lifetime cannot be explained by excitonic recombination, which is well below 1 ns for these GaAs QDs.

From the sharp Hanle peak for $X^-$, we therefore conclude that polarized resident electrons are produced in this experiment. Their lifetime can be understood as arising from the hyperfine interaction of the electron with a static nuclear environment [60]. Although this a measurement of a single dot, the measured time represents an "ensemble" dephasing, because the dot is measured over many optical cycles, during which nuclei will evolve. Furthermore, with nonresonant excitation, *e-h* pairs that recombine in one dot may have been influenced by polarized electrons in other parts of the sample.

The Hanle peak points downward in Fig. 4(c), unlike the usual Hanle effect in bulk semiconductors. This means that optical pumping enhances negative PL polarization. In general, the connection of negative polarization to optical pumping is not obvious, because the polarization of the trion is determined by its hole, not by the resident electron. Here, it can be understood with the simple mechanism shown in Fig. 10(b) [42]. As discussed in the previous section, dark excitons may be preferentially captured into the quantum dots, and the resulting trions are therefore negatively polarized. This process leads to optical pumping, because the spin down hole removes the original resident electron and leaves the photoelectron in its place.

InAs self-assembled QDs have stronger confinement potentials, and higher energy excited states of the dot can play a more prominent role in nonresonant optical pumping. As discussed earlier, a rich variety of exchange-induced relaxation processes is available among the trion excited states. We have shown in Fig. 6 how resonant absorption to those states can produce negatively-polarized PL. Although the particular relaxation path shown in Fig. 6 does not cause optical pumping, others paths do, and under nonresonant excitation, one can expect some optical pumping to occur. Measurements on ensembles of InAs QDs indeed support the connection between optical pumping and negative polarization.

*5.2 Resonant optical pumping*

Resonant optical pumping in QDs is analogous to the optical pumping originally demonstrated in atoms [54,55], where a QD is excited to a single discrete state, such as the lowest state of the trion. Although the process is quite simple to analyze and has a long history, it has only recently been demonstrated in QDs. Resonant pumping achieves very high fidelity, although because of its energy-specific nature, it may be difficult to use for pumping a large collection of dots simultaneously.

For a QD, the relevant energy levels are those shown in Fig. 10(c)—namely, the spin states of the resident electron and the trion. Pumping requires rapid and unidirectional transfer of population from one resident spin state to the other through excitation and emission from the trion state. Relaxation between the resident spins must be much slower than this optical process in order for population to build up in the new state. The experimental signature is depletion of the initial state, resulting in transparency of the optical transition. Very recently, three closely related examples of this process have been demonstrated [61,62,63]. Although the experimental technique is more or less the same, each approach highlights the importance of particular excitation and relaxation processes.

In the work of Atatüre et al [61], the sample was placed in a magnetic field parallel to the growth axis and to the propagation vector of the light (Faraday configuration). Circularly polarized laser light was tuned to a trion resonance, and the trion was excited with high probability. Although the diagonal recombination transition [Fig. 10(c)] was nominally forbidden by circular polarization selection rules, a small amount of light hole mixing in the trion opened up this channel, allowing recombination to the spin-down resident electron. Once trapped there, the QD was transparent to the excitation light, which differed in energy from the diagonal transition by the electron Zeeman energy. Only relaxation of the electron spin could return the population to the original state. This could occur through the nuclear hyperfine interaction or through co-tunneling with electrons in the doped substrate. The former was largely suppressed by the applied magnetic field, while the latter could be minimized by an appropriate choice of electric field.

Xu et al. reported a version of this process that improved upon the efficiency of the diagonal transition [62,64]. By placing the QD in a perpendicular magnetic field (Voigt configuration), the resident electron spin states are mixed, making the optical selection rules linear, so that diagonal transitions became optically bright. Using this approach, the pumping rate was increased by more than three orders of magnitude.  The proof of optical pumping (also in Refs [61,63]) is the lack of an absorption signal, except when a repumping transition (gray arrow) is simultaneously excited to transfer the population back to the initial state.  As with the Faraday geometry, this approach requires the suppression of the hyperfine interaction (magnetic field) and co-tunneling (applied bias).  A new caution also applies to this approach:  a small amount of near-resonant excitation with the same polarization is always possible on a re-pumping transition, due to the less restrictive selection rules.  This is suppressed by having a sufficiently large difference in Zeeman energies.

A third variation takes advantage of a resident hole spin [63].  The advantage is the insensitivity of the hole to the nuclear hyperfine interaction, which leads to spin lifetimes in the millisecond range and allows pumping in the absence of an applied magnetic field.  In fact, the hyperfine interaction plays an essential rather than detrimental role here, because it mixes the trion states (now distinguished by an unpaired electron spin) and allows efficient pumping to the opposite resident hole spin state.  With this method, all four optical transitions occur at the same energy, and so optical selection rules are particularly important to the fidelity of pumping.

*5.3 Spin Lifetimes*

The viability of QD spins for quantum information depends on their lifetimes being long enough to carry out a large number of operations.  Results of spin lifetime measurements in the last several years look very promising.  Energy relaxation times $T_1$ were measured to be ~20 ms for InAs QDs in a magnetic field of 4T [65].  The phonon-mediated relaxation process that governs this lifetime becomes less efficient at lower magnetic fields, and $T_1$ lifetimes around 1 second were measured in electrically-controlled GaAs QDs [66].  The loss of spin coherence, characterized by the time $T_2$, is the more important number for quantum information processing.  Typically, measurements on ensembles of QDs or on single QDs over a long period of time yield a much shorter dephasing time $T_2^*$ that is limited by ensemble fluctuations of the nuclear spin polarization through the hyperfine interaction [60,42].  However, recent phase-sychronization techniques have been used to overcome the inhomogeneous dephasing in InAs QDs, revealing a $T_2$ time of several microseconds [67].  This time is believed to be limited by the evolution of the nuclear spin environment through the hyperfine interaction [60].

## 6. Dynamic nuclear polarization

Bracker, et al., Fine structure and optical pumping of spins...A typical QD contains on the order of $10^5$ nuclei, and the hyperfine interaction between electron and nuclear spins strongly influences both. The hyperfine interaction in QDs has been a topic of considerable interest in recent years, because of its importance in electron spin lifetimes and optical orientation of nuclei. Here we focus on dynamic nuclear polarization (DNP), where a non-equilibrium polarization of electron spins causes the nuclei to become polarized. DNP was observed very early in single dot spectroscopy through Overhauser shifts in the work of Brown et al. [6]. A variety of kinetic processes are possible, and recent results on DNP of QD nuclei are rich and sometimes complex. Many of these phenomena can be better understood with the perspective of classic magnetic resonance experiments, so we consider some of these ideas throughout this section.

DNP occurs when the electron-nuclear spin system responds to being driven out of equilibrium, so we first describe the situation at equilibrium. The hyperfine interaction induces flip-flop transitions between electron and nuclear spins, conserving their total spin. In steady state, the populations $N_\mu$ of the QD nuclear spin sublevels ($\mu = -I,...,+I$) and $n_\uparrow$ ($n_\downarrow$) for up (down) electron spins [56] is

$$W_{\mu,\mu+1} N_\mu n_\uparrow = W_{\mu+1,\mu} N_{\mu+1} n_\downarrow, \qquad (4)$$

where $W_{\mu,\mu+1}$ is the probability of the flip-flop process $|\uparrow,\mu\rangle \to |\downarrow,\mu+1\rangle$, in which the projection of nuclear spin increasing by one and the projection of electron spin decreases by one. If the reservoir that supplies energy for flip-flop transitions is in thermodynamic equilibrium at temperature T, then the probabilities are related by

$$\frac{W_{\mu,\mu+1}}{W_{\mu+1,\mu}} = \exp\left(\frac{\Delta E}{kT}\right), \qquad (5)$$

where $\Delta E$ is the energy difference between the states $|\uparrow,\mu\rangle$ and $|\downarrow,\mu+1\rangle$ of the electron-nucleus system. The nuclear Zeeman energy is negligible, so $\Delta E$ reduces to the electron spin Zeeman energy $\Delta E = \mu_B g B_{ext}$. In equilibrium $(n_\downarrow/n_\uparrow)_{eq} = \exp(\mu_B g B_{ext}/kT)$, and Eq. 4 becomes $N_\mu = N_{\mu+1}$, i.e. the nuclei are practically unpolarized.

In order to achieve high nuclear polarization, it is necessary to destroy the equilibrium, hence the term "dynamic". From Eq. 4, which is general, this means that DNP requires the *n*'s or the *W*'s to vary from their equilibrium values. In the following sections we consider several DNP mechanisms that achieve this. These mechanisms are well-known in bulk semiconductors and can also be important in quantum dots.

*6.1 Overhauser effect*

In the Overhauser effect, the population distribution of electron spins does not correspond to the reservoir temperature T. In the classic Overhauser effect [68], electron spins are artificially depolarized ($n_\uparrow = n_\downarrow$) through absorption of microwaves or unpolarized light, and Eqs 4 and 5 give

$$N_{\mu+1}/N_\mu = \exp(\mu_B g B_{ext}/kT). \tag{6}$$

This ratio may be quite large, because the magnetic moment of the electron is ~1000 times larger than that of the nucleus.

With optical orientation, much larger nuclear polarization can be achieved, because $n_\downarrow$ and $n_\uparrow$ can be made substantially different. This works even at high temperature and low field, giving $N_{\mu+1}/N_\mu = n_\uparrow/n_\downarrow$. In this case, there is a simple relation between electron and nuclear polarizations, when the polarizations are not very large and in the absence of leakage [20]:

$$P_N = \frac{(I+1)}{(S+1)} P_S. \tag{7}$$

One can see that the sign of DNP is determined by the sign of the photoelectron polarization. This has been reported for neutral QDs by various laboratories. A negatively charged QD plays a role similar to a donor-bound electron, so its optical orientation will also induce nuclear polarization, as demonstrated first by Dzhioev et al. [69] in an ensemble of InP quantum dots. In some systems, the resident electron can be polarized oppositely from the photoelectron, because of trion relaxation induced by electron co-tunneling with a nearby reservoir. In this case, the nuclei will also be polarized in the opposite direction. [58,59]

*6.2 The Solid Effect*

*6.2.1 Classical Solid Effect.* The solid effect is another DNP mechanism, proposed by Abragam [70,71]. It is not as well known as the Overhauser effect but has been extensively studied in bulk materials by ESR-NMR techniques. The name "solid effect" comes from its importance in the study of paramagnetic centers in solids, where a localized electron interacts with nuclei in a limited region, in contrast to the situation for liquids, where much NMR research was done. We

Bracker, et al., Fine structure and optical pumping of spins…

consider it here, because it can lead to interesting new phenomena for electrons localized in nanostructures, and it is likely to generate new challenges and opportunities for researchers. Indeed, we argue that it already plays an important role in some recent observations of nuclear polarization in QDs.

The idea of the solid effect is very simple. In the version considered by Abragam, the energy required for a flip-flop transition comes from an applied radio frequency field rather than a thermal reservoir. An rf field of the appropriate frequency can induce weakly allowed flip-flop stimulated transitions between electron and nuclear spins. In this case, the DNP is still described by Eq. 4, but instead of Eq. 5, we have $W_{\mu,\mu+1} = W_{\mu+1,\mu}$. The DNP takes place even if the electron spins remain in equilibrium (short spin relaxation time), and we get from Eq. 4

$$N_{\mu+1}/N_\mu = (n_\uparrow/n_\downarrow)_{eq} = \exp(-\mu_B g B_{ext}/kT). \tag{8}$$

Comparing Eqs (6) and (8), we see that this DNP mechanism is just as effective as the classical Overhauser effect but produces nuclear polarization with the *opposite* sign.

Recently, Laird et al. demonstrated a version of the solid effect in a GaAs double-dot. [72]. The hyperfine interaction was modulated via an electric dipole spin resonance, thereby inducing flip-flop transitions, and the nuclear polarization had the opposite sign from that expected for the Overhauser effect. Rudner et al. [73] provided an explanation, noting that this was not a case of the specific version of the solid effect described by Abragam, where the r.f. field induces a forbidden *e-n* transition. Nevertheless, it is an example of the solid effect, except that the rate constants *W* are equalized by a different interaction, and Eq. 8 applies.

*6.2.2. Optical solid effect.* More generally, the solid effect can occur whenever the rate constants $W_{\mu,\mu+1}$ and $W_{\mu+1,\mu}$ are forced to have a non-thermodynamic relation to each other. For example, in the optical range, not only stimulated but also spontaneous transitions are important, so the *W*'s can be substantially different. DNP is then possible even at high temperature, where electrons are unpolarized ($n_\downarrow = n_\uparrow$) and none of the classic DNP effects are efficient. In this case, the nuclei are polarized, with

$$\frac{N_{\mu+1}}{N_\mu} = \frac{W_{\mu,\mu+1}}{W_{\mu+1,\mu}} \neq 1, \tag{9}$$

This ratio is weakly temperature dependent. The optical solid effect has been observed for the singlet-triplet excitons in bulk silicon [74]. There, the singlet state was empty due to strong spontaneous emission, whereas the two split-off ±1 triplet states were populated because their optical decay was forbidden. A magnetic field removed the degeneracy of the triplets, bringing one closer to the singlet than the other, resulting in DNP by Eq. 9.

A neutral exciton in a QD, excited by unpolarized light, is an important candidate for the high temperature solid effect. The hyperfine interaction induces transitions between bright and dark states: both $\Downarrow\uparrow \leftrightarrow \Downarrow\downarrow$ and $\Uparrow\downarrow \leftrightarrow \Uparrow\uparrow$. In this process, the electron-nuclear flip-flop converts a dark state into a virtual bright state, which emits a photon [41]. Bright excitons disappear rapidly through recombination, so that $n_{\pm 1} \ll n_2 = n_{-2}$, and only the flip-flop transitions $\Downarrow\downarrow; \mu+1 \rightarrow \Downarrow\uparrow; \mu$ and $\Uparrow\uparrow; \mu \rightarrow \Uparrow\downarrow; \mu+1$ take place. The nuclear spin decreases in the former case and increases in the latter case, and there is no DNP because the probabilities are equal. However, in an applied magnetic field in the Faraday geometry, the two processes are not equivalent: the energy difference $(\delta_0 + \mu_B g_e B_{ext})$ between the states $\Downarrow\downarrow$ and $\Downarrow\uparrow$ increases, while the difference $(\delta_0 - \mu_B g_e B_{ext})$ between $\Uparrow\uparrow$ and $\Uparrow\downarrow$ decreases by the electron Zeeman energy. For small magnetic field $(\mu_B g_e B_{ext} \ll \delta_0)$, we have:

$$W_{\mu+1,\mu} \propto \frac{1}{(\delta_0 + \mu_B g_e B_{ext})^2}; \qquad W_{\mu,\mu+1} \propto \frac{1}{(\delta_0 - \mu_B g_e B_{ext})^2}. \qquad (10)$$

It follows from Eqs 9 and 10 that $\frac{N_{\mu+1}}{N_\mu} \approx 1 + 4\frac{\mu_B g B_{ext}}{\delta_0}$ and the QD nuclear polarization is

$$P_N = \frac{1}{I}\sum_{\mu=-I}^{+I} \mu N_\mu \approx \frac{4}{3}(I+1)\frac{\mu_B g_e B_{ext}}{\delta_0}. \qquad (11)$$

Unlike the classical Overhauser and solid effects, the splitting $\delta_0$ enters in the denominator instead of temperature. At high temperature, this DNP mechanism can be more effective than the classical effects, because $\delta_0 \ll T$. To our knowledge, it has not yet been identified in QDs, however it should persist to a measurable degree even in low temperature experiments such as those of Ref. [41], when linearly polarized light is used.

Merkulov [75] treated the exciton-nuclear spin system excited by circularly polarized light, a scenario in which both the optical Overhauser effect, $(n_2 \neq n_{-2})$ and the solid effect $(W_{\mu,\mu+1} \neq W_{\mu+1,\mu})$ must be considered. He found both self-polarization and a striking bistability in the spin system. These topics are considered in Section 6.4.

*6.3. Nuclear spin "leakage" and dynamic nuclear polarization*

Nuclear spin relaxation (leakage) prevents DNP from producing 100% nuclear polarization. Eq. 4 neglects possible sources of leakage. The highest polarizations observed to date are around 60% [41,42]. Here we describe a few important sources of leakage and ways to avoid them.

Leakage sources can be categorized as intrinsic and extrinsic. Extrinsic sources include paramagnetic impurities, charge carriers in the matrix, etc. The intrinsic sources are inherent to the electron-nuclear spin system itself. They include interactions that do not conserve total spin, such as the anisotropic part of the hyperfine interaction and the nuclear dipole-dipole interaction. The latter is the most important for shallow donor electrons and likewise for QDs.

The nuclear dipole-dipole interaction prevents DNP at zero magnetic field [56]. An external magnetic field, $B_{ext} \gg B_L$, (i.e. larger than the local field $B_L \sim 1G$, created by the surrounding nuclei) suppresses these dipole-dipole transitions and restores DNP. The effect is well-known in bulk GaAs and has also been observed in individual neutral GaAs QDs [41]. It is observed as a dip in the Overhauser shift as a function of magnetic field around $B_{ext}$=0, with a width of ~160 mT in the QDs [Fig. 8(d)]. The dip is hundreds of times broader than in bulk GaAs. This dramatic difference has been explained as a suppression of DNP by the electron-hole exchange interaction in QD excitons [41]. The rate $1/T_{1e}$ for DNP is slowed by the large energy mismatch between spin flips of nuclei and electrons, with electron spin flips occurring through the conversion between dark and bright excitons (e-h exchange energy, $\delta_0$). Furthermore, the rate of nuclear depolarization, i.e. leakage ($1/T_L$), is enhanced compared to the bulk. The nuclear dipole-dipole interaction weakly mixes the different nuclear spin projection states, on the order of $B_L/B_{ext}$. These transitions are induced by the hyperfine field of the electrons, acting on the nucleus, during the long lifetime of the dark exciton ($\tau_d \gg \tau_b$). Both effects combine to increase the width of the dip in nuclear polarization as a function of magnetic field by the factor $\delta_0 \sqrt{\tau_d \tau_b} / \hbar$, which is ~300 for the GaAs QDs.

In charged QDs, the situation is quite different, with nuclear polarization observed at $B_{ext}$=0. This suppression of the dipole-dipole interactions in InAs QDs was recently explained as a result of the Knight field imposed on the nuclei by optically oriented electrons [58]. However, at least for InP dots, a different explanation is required [76]. There, the Knight shift explanation is inconsistent with an observed broadening of the Hanle curve. Instead, it is argued that a strain-induced nuclear quadrupole interaction suppresses the nuclear spin flip and makes zero-field DNP possible.

Bracker, et al., Fine structure and optical pumping of spins...

Even if an external magnetic field is applied, there is still leakage related with the secular part of the dipole-dipole interaction. Flip-flop transitions between neighboring nuclear spins are still possible and induce nuclear spin diffusion [77,78] out of QD region into the matrix with a characteristic time $T_L \sim R^2/D_N$. For a QD with diameter 10 nm and diffusion coefficient $D=10^{-13}$ cm$^2$/s one gets $T_L \sim 10$ sec. This is comparable with DNP times $T_{1e} \sim W^{-1} = 1-10$ sec.

There are a few possibilities to overcome nuclear spin diffusion: (i) pump for a long time to polarize nuclei in the nearby matrix [77,78]; (ii) a non-uniform Knight field of spin-polarized electrons blocks diffusion [79]; (iii) non-uniform stress in a self-organized QD produces a non-uniform quadrupole interaction that inhibits spin transfer [80]; (iv) if the QD contains a pure nuclear isotope while the barrier consists of a different isotope, then an external field ~20 G should block diffusion [81].

*6.4. Non-linear properties of the electron-nuclear system in a QD*

*6.4.1. Origin of non-linearity*. It was recognized long ago that the electron-nuclear spin system is essentially nonlinear [82]. The main source of nonlinearity in DNP is the effective magnetic field $\vec{B}_N \sim \vec{I}_N$ of polarized nuclei [83], which plays a twofold role: (i) and (ii) We now consider these in turn.

(i) $B_N$ changes the level splitting, thus affecting the rates of electron spin flip and electron-nuclear flip-flop transitions. It follows from Eq. 7 in the absence of leakage that the nuclear spin is determined solely by the mean spin of resident (donor bound) electrons, due to the Overhauser effect (Section 6.1). The magnetic field $B_N$ of polarized nuclei influences the relaxation rate of the electron spin thus making $S(B_N)$ dependent on the nuclear spin $I_N$. Eq. 7 becomes transcendental, allowing for multiple solutions [84]. Such a feedback effect for excitonic spin relaxation has been reported for InP QDs by Dhzioev et al. [85]

Even if the mean spin $S$ is not affected by $B_N$, another important nonlinearity arises due to leakage with characteristic time $T_L$. The right hand side of Eq. 7 should include a leakage factor $f_N = \dfrac{T_L}{T_L + T_{1e}}$ [20]. The DNP time $T_{1e}$ of flip-flop transitions depends on the total field $B_{ext} + B_N$ [84], which allows for multiple solutions. Competition between DNP and leakage manifests itself in non-trivial behavior of the electron-nuclear spin system in a QD. A pronounced hysteresis in nuclear polarization for X$^-$ and X$^+$ in InAs QDs has been measured as a function of magnetic field and laser intensity [86,87,88]. Similar nonlinear physics appears in the presence of the solid effect. Merkulov [75] considered the exciton-nuclear spin system under the action of a large $B_N$ field, so that the probabilities of forward and reverse flip-flop transitions are different. He found a striking hysteresis behavior originating from bistability. The hysteresis for the $X^0$ in a QD has been observed by Tartakovskii et al. [89], although their analysis did not include electron-hole exchange, which is comparable with the hyperfine interaction and cannot be neglected. Further efforts are necessary to illuminate the physics of the observed hysteresis.

(ii) $B_N$ affects the direction of the electron spin by means of spin torque. If electron and nuclear mean spins are not collinear (for example, in an oblique external magnetic field), then a spin torque $\vec{S} \times \vec{B}_N$ acts on electrons, which should be inserted into the Bloch equation for spin $\vec{S}$. The value and direction of $\vec{B}_N$ are determined in turn by $\vec{S}$. Note that this takes place even in the absence of the effect of $\vec{B}_N$ on relaxation and polarization rates. Under these conditions, not only bistabilty, but also oscillations of polarization, are possible. These types of nonlinearities have been studied in detail in bulk semiconductors [82], although so far there are no reports on this subject in QDs.

*6.4.2. Dynamic nuclear self-polarization.* Perhaps the most spectacular manifestation of DNP nonlinearity is spontaneous nuclear polarization under non-polarized excitation, i.e. $\vec{S} = 0$. The well-known Dyakonov-Perel mechanism for self-polarization [56] is based on the classical Overhauser effect and the thermodynamic relation of Eq. 5, where $\Delta E = AIP_N$. This mechanism requires very low temperature $T \leq T_c \equiv AI(I+1)/3 \sim 1K$.

Korenev [90] proposed a spontaneous self-polarization mechanism based on the solid effect that can occur at higher temperatures. The effective magnetic field of polarized nuclei splits the electron spin levels by the value $AIP_N$. This term replaces the electron Zeeman energy in Eq. 11, and a non-trivial solution develops when the electron-hole exchange interaction $\delta_0 \leq 4AI(I+1)/3$. For typical values of $\delta_0 \approx A \approx 100\,\mu eV$, this condition can be satisfied for any spin $I > 1/2$. This scenario is weakly temperature dependent. Korenev's proposal is based on the accumulation of dark excitons (bright excitons recombine), and the convergence of one pair of bright and dark exciton levels (e.g. ⇑↑ and ⇑↓), which accelerates DNP. This general approach, involving shifts of electron spin levels and feedback to DNP, was Bracker, et al., Fine structure and optical pumping of spins...

also exploited by Rudner et al. [91] to predict self-polarization in electronic transport through the double-dot system.

Recently, Korenev [92] considered self-polarization under quasi-resonant excitation of a QD charged with a single electron. Again, this is based on the solid effect, but unlike previous versions, no level convergence takes place: the spontaneous magnetic field of spin-polarized nuclei shifts the *optical transition* energy toward resonance with the photon energy.

Self-polarization remains a substantial scientific opportunity for experimentalists.

## 7. Conclusion

Single dot spectroscopy was born out of the need to overcome inhomogeneous broadening of QD ensembles and to resolve individual quantum states. With the resolution it provides, the atom-like nature of QDs has become explicitly clear. At the same time, a variety of phenomena have been uncovered that are particular to semiconductor quantum dots and are not seen in atoms. In the future, the uniformity of QDs may never improve so much that high resolution spectroscopy of spins is practical in ensembles. However, this can be seen as a reflection of one of their great advantages, namely, the ability to engineer QD properties over a wide range. Some of the applications that we now imagine for QDs, including quantum information processing with spins, could make use of a collection of inequivalent discrete elements. For such an approach, we need dot-specific ways to interact with that collection, and the tools of single dot spectroscopy are a natural step in that direction.

## Acknowledgement

This work was supported by NSA/ARO and ONR.

Bracker, et al., Fine structure and optical pumping of spins...

Bracker, et al., Fine structure and optical pumping of spins...

Figure 1. a) Ensemble photoluminescence spectrum of InAs self-assembled quantum dots. Higher shell structure (p,d,f) is populated under high intensity laser excitation. b) Band edge profile of a typical Schottky diode used for single dot spectroscopy in this work. c) Bias-dependent PL map of a single InAs quantum dot. Shell structure above the s-shell is populated when extra electrons are electrically injected. d) Close-up of PL map for s-shell exciton structure in another QD. Excitons (X) and biexcitons (2X) with various charges are shown. $X^{2-}$ and $X^{3-}$ transitions show spin fine structure.

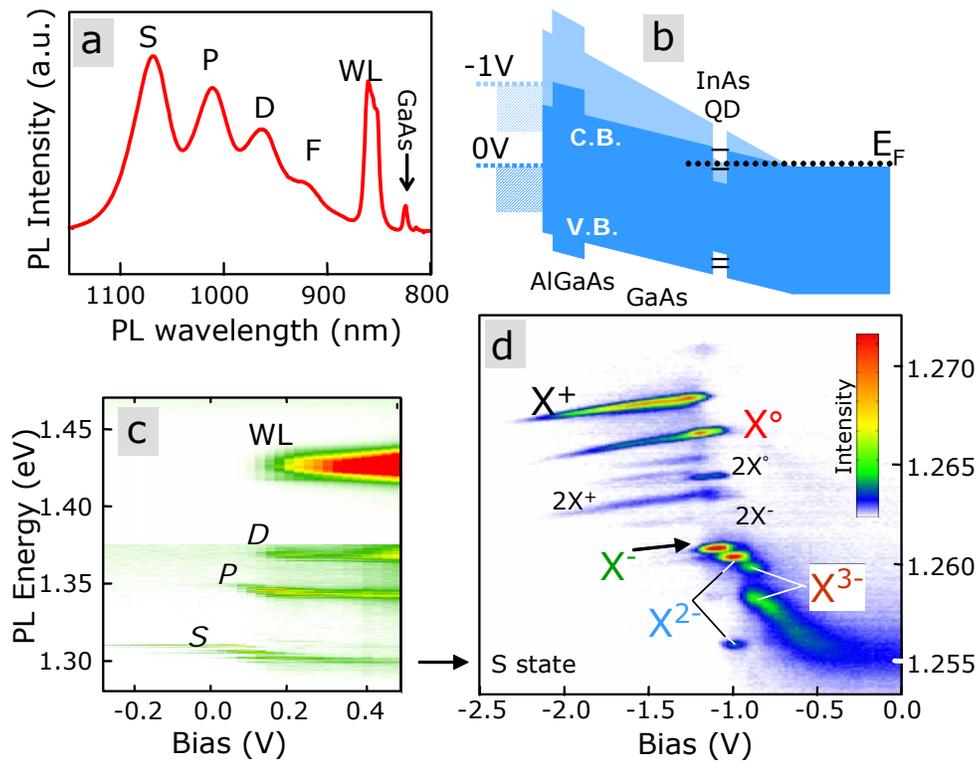

Figure 2. a) Photoluminescence spectra of GaAs/AlGaAs quantum wells with several widths (indicated on the right of each spectrum). Red lines trace neutral exciton $X^0$ and positive trion $X^+$ peaks. The two pairs of red lines correspond to two well widths, varying by one atomic layer of GaAs. The top spectrum shows $X^0$ and $X^+$ peaks from an individual interface QD in a 2.8 nm quantum well [29]. Higher energy peaks come from different dots. b) Band edge profile of a Schottky diode heterostructure used in spectroscopy of individual GaAs interface QDs. c) Bias-dependent PL map of a single GaAs quantum dot. d) PL polarization of the lines in (c) [42]. $X^0$ is unpolarized due to anisotropic exchange, while $X^+$ and $X^-$ show the polarization of the unpaired electron and hole spins, respectively.

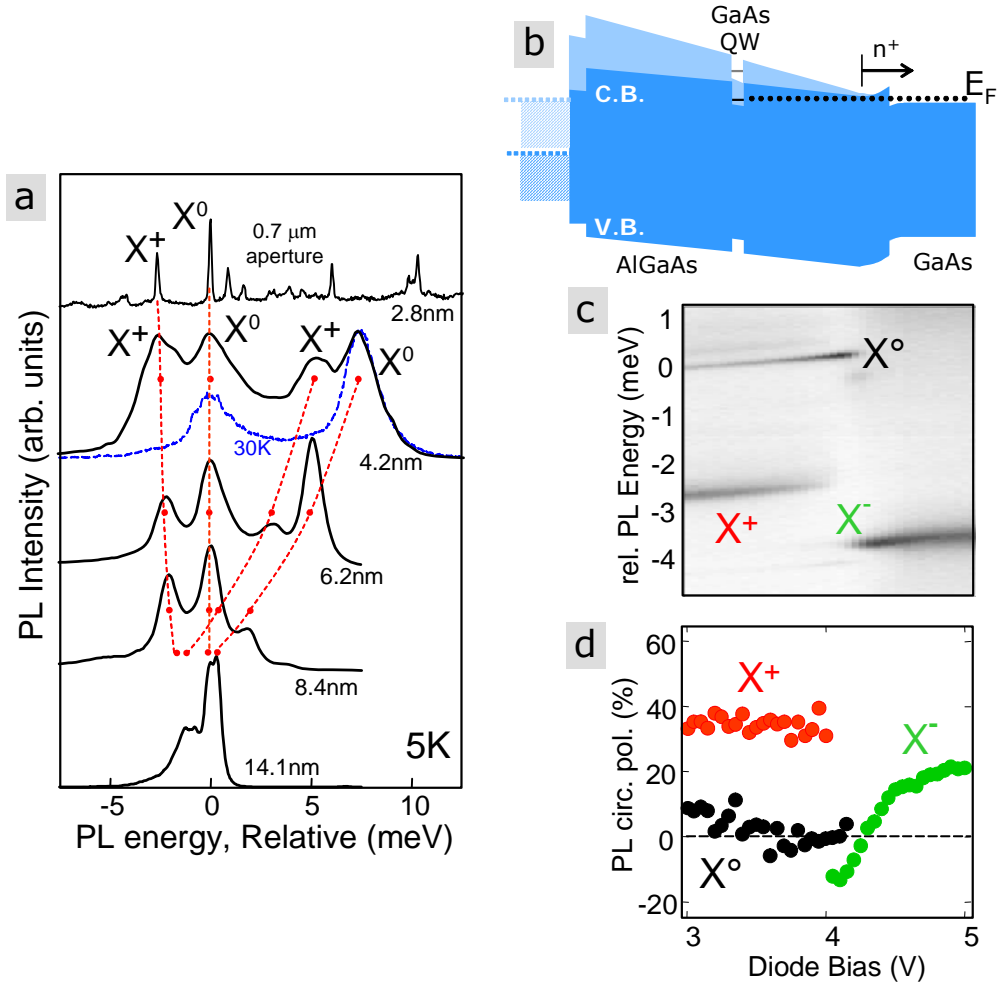

Bracker, et al., Fine structure and optical pumping of spins...

Figure 3. a) Energy level structure of the neutral exciton. Linearly polarized spin wavefunctions existing at zero magnetic field are shown (Notation: $\Uparrow, \Downarrow \equiv$ heavy holes with $m=\pm 3/2; \uparrow, \downarrow \equiv$ electrons with $m=\pm 1/2$). Electron-hole exchange energies $\delta_0$ and $\delta_b$ are indicated. Transition polarizations are x or y for B=0 and $\sigma+$ or $\sigma-$ for B$\neq$0. b) Schematic of neutral exciton energy levels in a magnetic field, with circularly polarized spin wavefunctions. Exciton fine structure of the upper "bright" states can be probed by measuring the change in PL circular polarization as a function of magnetic field [45,31,46].

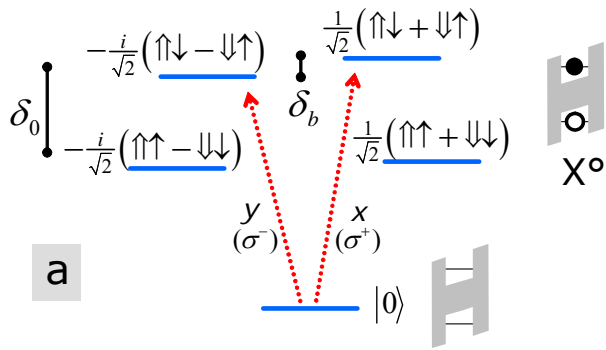

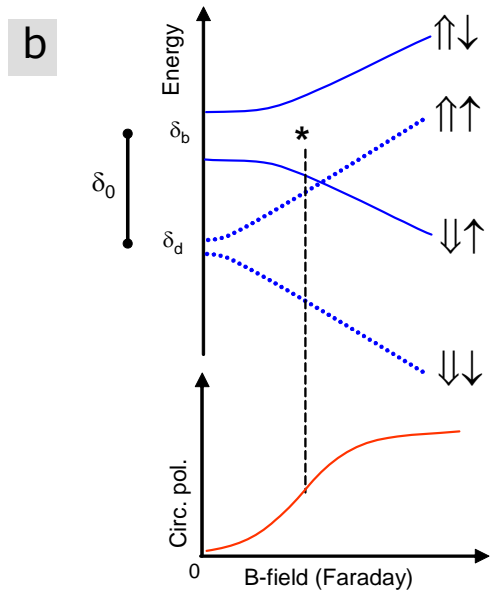

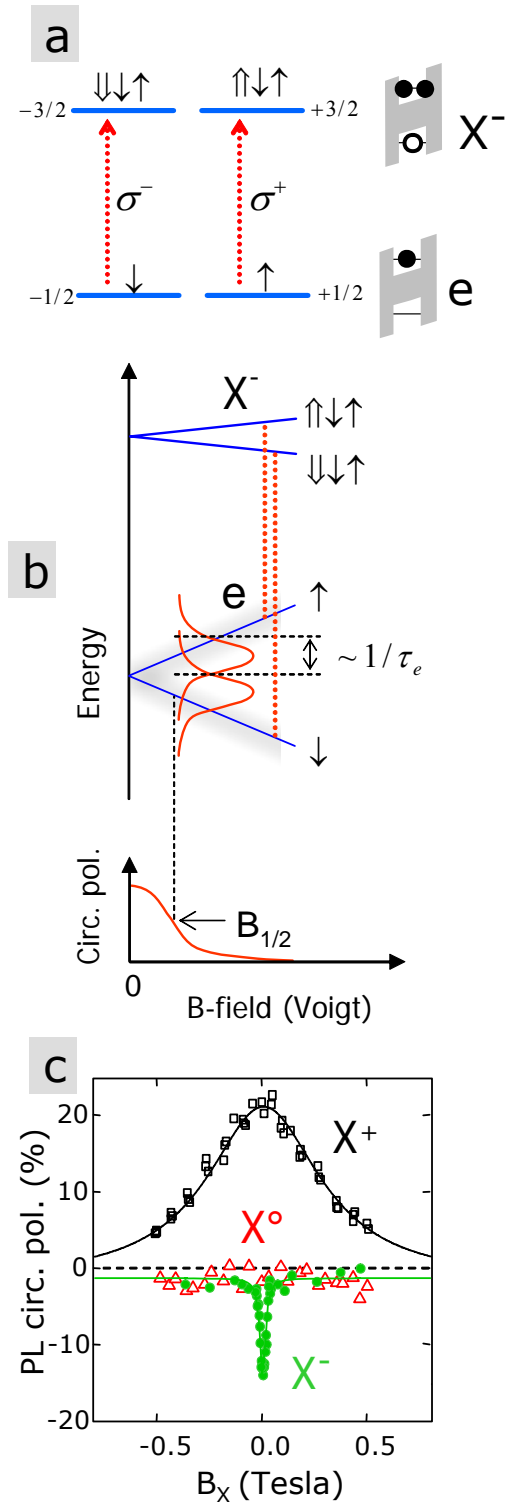

Figure 4 a) Energy levels of negatively charged exciton ground singlet state and resident electron, labeled with spin wavefunctions and angular momentum projection. (Asymmetrized singlet wavefunctions are indicated in Fig. 6.) Circularly polarized allowed transitions are shown.  b) Schematic depiction of the Hanle effect as a zero field level crossing for an electron spin.  Electron spins are depolarized when the Zeeman splitting becomes comparable to the inverse spin lifetime.  c) Hanle effect measurement for a negative trion, positive trion, and neutral exciton in an individual GaAs QD [42].  The broad $X^+$ feature results from the recombination-limited lifetime of the electron in $X^+$, while the $X^-$ feature probes the longer-lived resident electron.

Bracker, et al., Fine structure and optical pumping of spins...

Figure 5 a) Photoluminescence spectra of a neutral exciton and positive trion in a GaAs QD with magnetic field (6T) at three angles. Dark state transitions of the exciton (m=±2) and dark diagonal transitions of the trion become allowed away from θ=0° (Faraday geometry). b) Full energy dependence of exciton and trion fine structure as a function of the angle of the magnetic field. Note: in Ref. [29], from which these graphs were taken, the trion was described as negatively charged. Later work [16] showed that it was actually a positive trion.

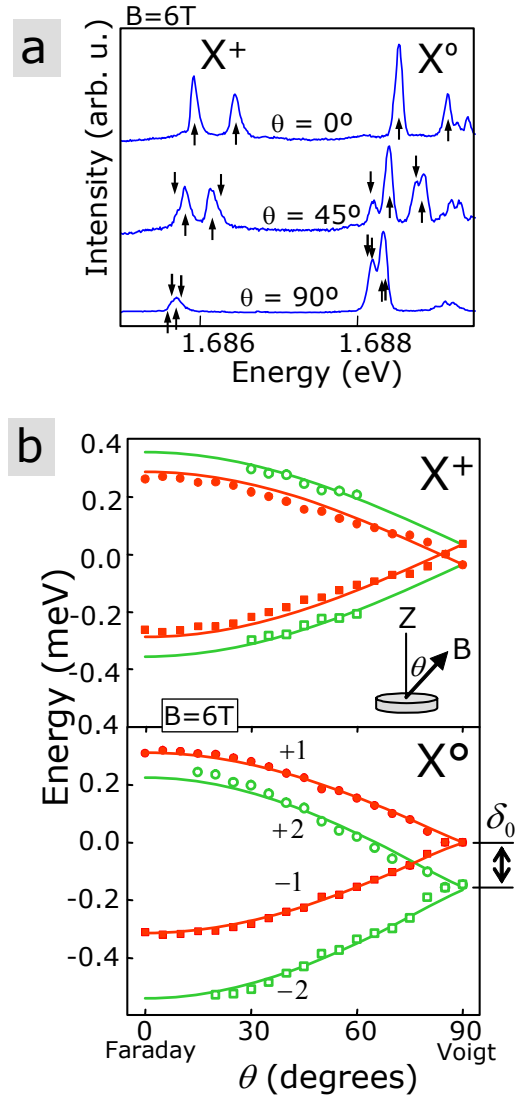

Fig. 6 a) Spin fine structure of a negative trion in an InAs QD, including excited states when one electron is in a p-like QD orbital (from Ref. [34]). Axially symmetric exchange splittings (Δ) and asymmetric interactions (δ, not splitting) are indicated. Black wavy arrows in part (a) indicate photoexcitation of doublet structure in part (c). Colored arrows represent recombination pathways. The higher energy triplet excitation results in negative photoluminescence polarization. Large and small spin arrow symbols correspond to "s" and "p" orbital states of the QD b) Broad view of PL excitation spectrum for two PL detection polarizations. The arrow indicates the polarized doublet. c) Close up of polarized PLE doublet.

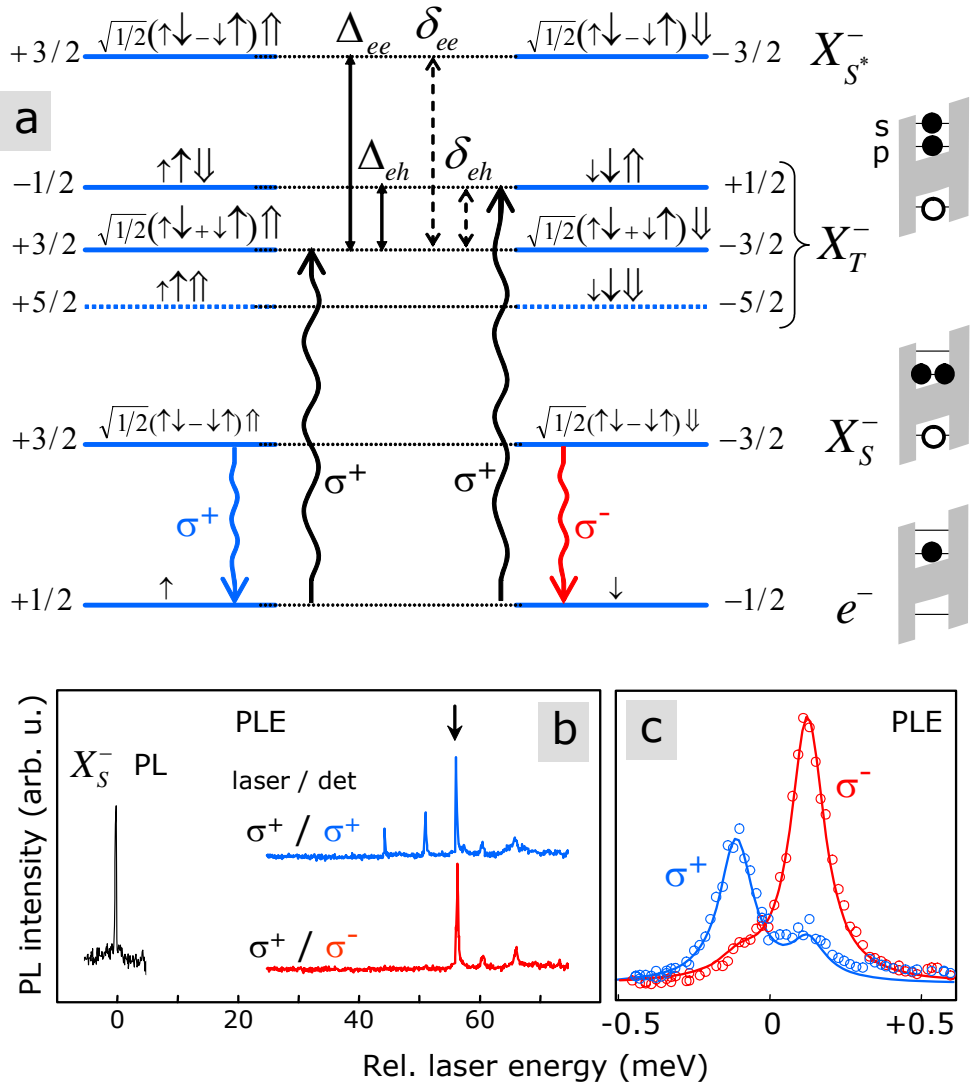

Bracker, et al., Fine structure and optical pumping of spins...

Fig. 7 a) Cross sectional STM image of an InAs quantum dot molecule. b) Schematic depiction of the anticrossing of exciton energy levels where interdot and intradot levels meet. Orbitals with bonding and antibonding character form at the anticrossing. c) Electric field-dependent photoluminescence map (and simulation) of spin fine structure [37] of a neutral exciton in a quantum dot molecule with an anticrossing resulting from hole tunneling. The PL map was measured at transverse magnetic field $B_x$=5T to make "dark" transitions visible. The anisotropic *e-h* exchange splitting ($\delta_0$) of the spatially direct exciton is indicated. d) PL map and simulation for a positive trion in a quantum dot molecule. Singlet and triplet eigenstates are indicated in the simulation. The nominal triplet with zero spin projection ($t_0$) is mixed with singlets. Dark triplets ($t_\pm$) are shown by dotted lines. Anisotropic *e-h* exchange in the left dot is observed for the spatially indirect transition.

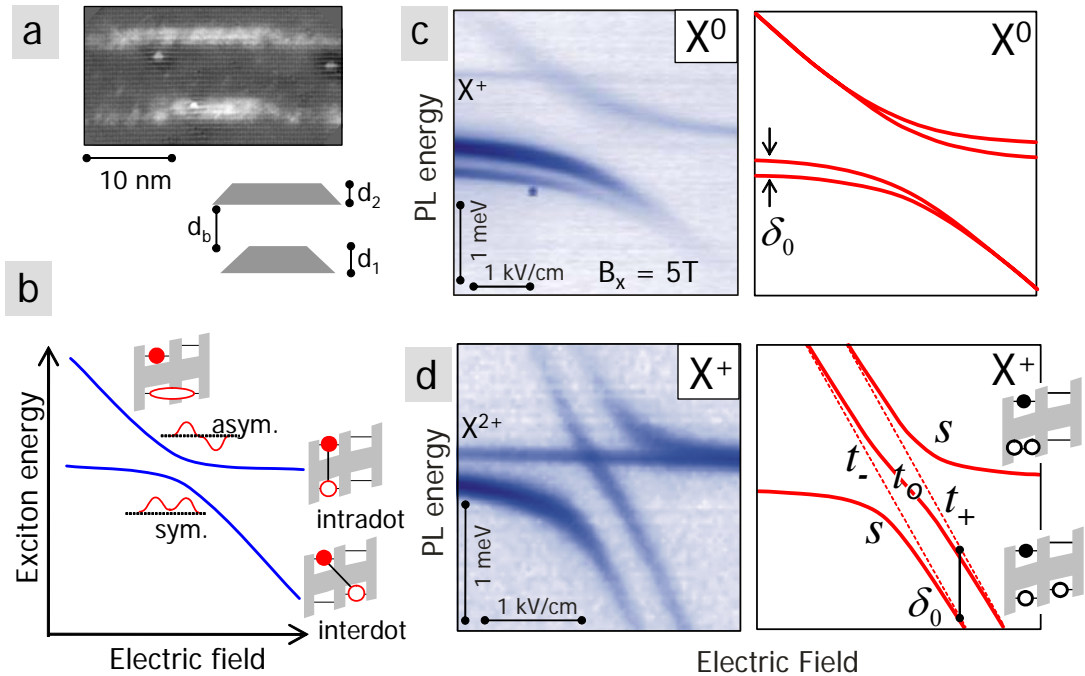

Bracker, et al., Fine structure and optical pumping of spins...

Fig. 8 a) Photoluminescence spectra of a Zeeman doublet for a neutral GaAs QD exciton at several longitudinal magnetic fields. The Overhauser effect causes the splittings to be asymmetric about $B_{ext}=0$. b) Energies of peaks in (a) as a function of external magnetic field ($\sigma^+$ laser excitation). Bright states and dark states (simulation only) are shown. c) Diamagnetic shifts extracted from (b) and from analogous data for $\sigma^-$ laser polarization. d) Zeeman splittings extracted from (b) and from analogous $\sigma^-$ data. The energy difference between the data for $\sigma^+$ and $\sigma^-$ is the result of the Overhauser effect. Graphs are taken from Ref. [41].

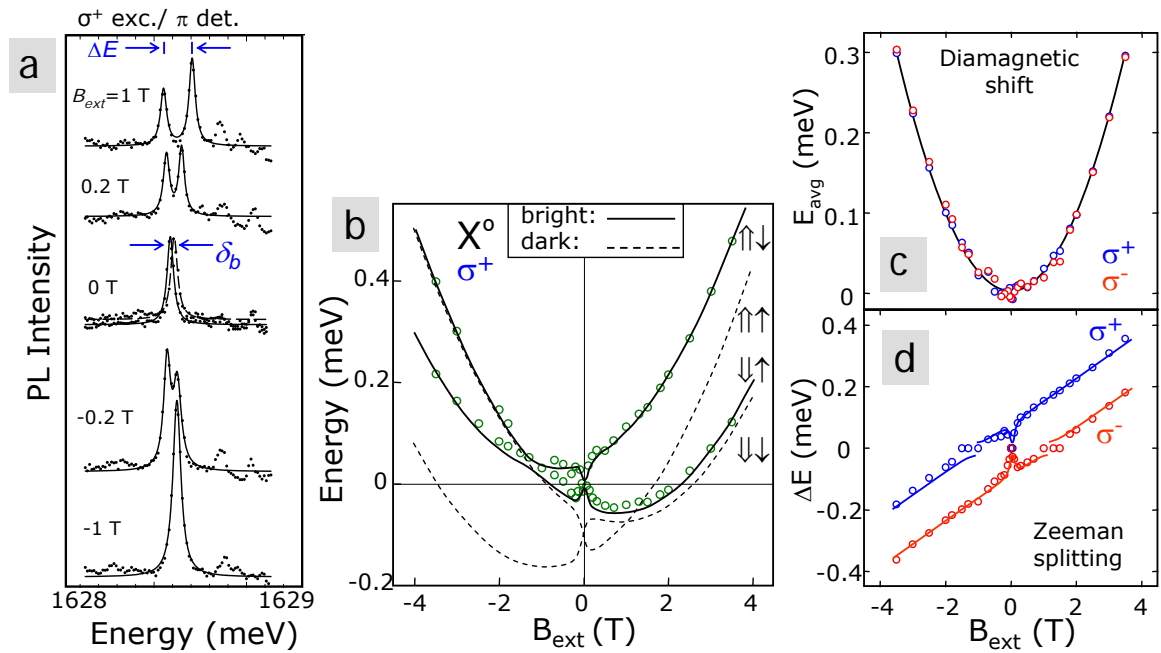

Bracker, et al., Fine structure and optical pumping of spins...

Fig. 9 a) Exciton photoluminescence peak for a GaAs QD in a magnetic field, showing the effect of nuclear magnetic resonance on Overhauser shifts. The peak splitting becomes independent of laser polarization when radiofrequency radiation is turned on, because dynamic nuclear polarization is eliminated. b) NMR resonances for Ga and As nuclei measured as a change in the spin splitting caused by nuclear depolarization. Taken from Refs [44, 43]

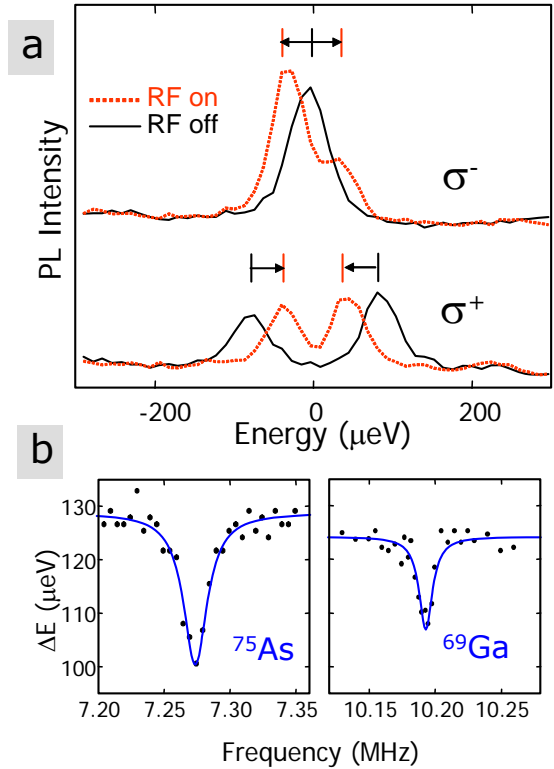

Bracker, et al., Fine structure and optical pumping of spins...

Fig. 10 a) Energy levels for non-resonant optical pumping of a resident electron spin in a quantum dot. Optical pumping results when relaxation follows $R_2$. b) Schematic of dark exciton mechanism for optical pumping and negative PL polarization. Dark excitons (dk) can only form trions, while bright excitons (br) can also recombine. The recombination channel for bright excitons reduces their population prior to capture and leads to negative polarization of trions. The original resident electron is shown in black and is replaced only for the dark exciton channel. (c) Generalized schematic for resonant optical pumping. The diagonal recombination channel transfers population to the spin-down electron state.

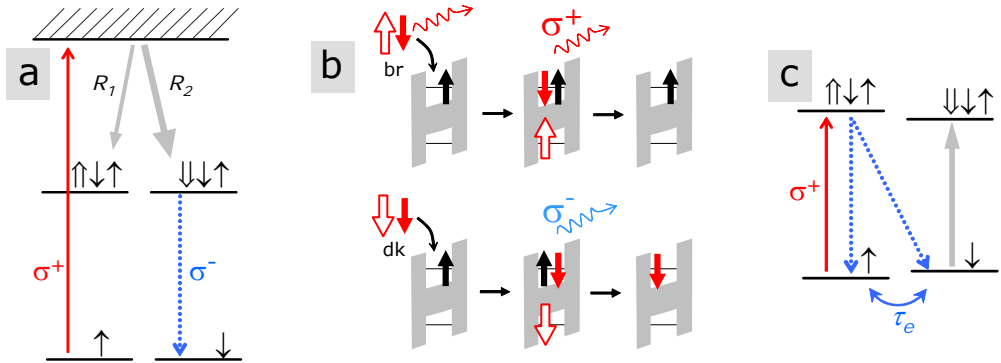